\documentclass[conference]{IEEEtran}

\usepackage{bm}
\usepackage{graphicx}
\usepackage{url}
\usepackage{caption}

\usepackage{balance}

\begin{document}

\title{Extremely scalable algorithm \\
for 10$^8$-atom quantum material simulation \\
on the full system of the K computer
}


\author{\IEEEauthorblockN{Takeo Hoshi and Hiroto Imachi}
\IEEEauthorblockA{Department of Applied Mathematics and Physics\\
Tottori University \\
4-101 Koyama-Minami, Tottori, Japan\\
Email(Hoshi): hoshi@damp.tottori-u.ac.jp}
\and
\IEEEauthorblockN{Kiyoshi Kumahata, Masaaki Terai, \\ Kengo Miyamoto, Kazuo Minami and
Fumiyoshi Shoji}
\IEEEauthorblockA{Operations and Computer Technologies Division \\
RIKEN Advanced Institute for Computational Science \\
7-1-26 Minatojima-minami-machi, Chuo-ku, Kobe, Japan \\
}
}

\maketitle

\begin{abstract}
An extremely scalable linear-algebraic algorithm 
was developed for quantum material simulation
(electronic state calculation)
with 10$^8$ atoms or 100-nm-scale materials.
The mathematical foundation is generalized shifted linear equations 
($(zB - A) \bm{x} = \bm{b}$), 
instead of conventional generalized eigenvalue equations.
The method has a highly parallelizable mathematical structure.
The benchmark shows
an extreme strong scaling 
and a qualified time-to-solution 
on the full system of the K computer.
The method was demonstrated 
in a real material research for ultra-flexible (organic) devices, 
key devices of next-generation Internet-of-Things (IoT) products. 
The present paper shows that 
an innovative scalable algorithm for a real research
can appear by the co-design 
among application, algorithm and architecture.
\end{abstract}

\IEEEpeerreviewmaketitle



\section{Introduction}

Large-scale quantum material simulation (electronic state calculation)
is a major field of computational science and engineering. 
Calculations for one-hundred-million (10$^8$) atoms or 
100-nano-meter(nm)-scale systems
have a strong need for innovative industrial products but
are far beyond the computational limit of the present standard methods.
The present paper reports that 
a novel linear algebraic algorithm 
\cite{HOSHI-MARNOLDI, HOSHI-2013-JPSJ, 
HOSHI-2014-JPSCP, HOSHI-2014-POS, IMACHI-2016-EIGENKERNEL,HOSHI-GREECE2}
shows an extreme strong scaling and 
a qualified time-to-solution on the full system on the K computer
with 10$^8$ atoms or 100-nm-scale systems.
The algorithm was implemented in our code ELSES
(=Extra-Large-Scale Electronic Structure calculation;
\url{http://www.elses.jp/}).
The method was demonstrated with condensed polymer systems
that 	appears in an academic-industrial collaboration research
for next-generation Internet-of-Things (IoT) devices.

The present paper is organized as follows;
The background or the algorithm
is presented 
in Sec.~\ref{SEC-BACKGROUND} or  Sec.~\ref{SEC-ALGO}, respectively. 
The benchmark and their analysis are given in Sec.~\ref{SEC-BENCH}.
The application in real research is given in Sec.~\ref{SEC-SCIENCE}.
The conclusion is given in Sec.~\ref{SEC-CONCLUSION}.

\section{Background \label{SEC-BACKGROUND}}

\subsection{Large-scale eigenvalue problem and its difficulty \label{SEC-COMP-PROBLEM}}

A mathematical foundation of electronic state calculations 
is a generalized eigenvalue problem of
\begin{eqnarray}
A  \bm{y}_k = \lambda_k B \bm{y}_k.
 \label{EQ-GEV-EQ}
\end{eqnarray}
The matrices $A$ and $B$ are Hamiltonian and the overlap matrices, respectively.
These matrices are $M \times M$ Hermitian matrices and  $B$ is positive definite.
In this paper, these matrices are real-symmetric.
An eigenvalue ($\lambda_k$) or eigenvector ($\bm{y}_k$) represents the energy or quantum wavefunction $\phi_k(\bm{r})$ of one electron, respectively.
In typical cases, the matrix size $M$ is 
proportional to the number of atoms $N$ ($M \propto N$).

Direct eigenvalue solvers consume ${\rm O}(M^3)$ operation costs and
their practical limit  is the matrix size of $M=10^6$ for the current supercomputers.
Recently, a million dimensional eigenvalue problem, the largest problem as far as we know,
was solved by an optimally hybrid solver 
(EigenKernel; \url{https://github.com/eigenkernel/})
\cite{IMACHI-2016-EIGENKERNEL}
with the two modern solvers of ELPA \cite{ELPA} and EigenExa \cite{EIGENEXA}.
The ELPA routine was used for the reducing procedure 
from the generalized eigenvalue problem into the standard one,
while the EigenExa routine was used so as to solve the reduced standard problem. 
The elapsed time on the K computer is $T_{\rm elaps}=9,939$ sec 
with $n_{\rm node}=41,472$ nodes 
and $T_{\rm elaps}=5,516$ sec on the full system (with $n_{\rm node}=82,944$ nodes).

The large-scale problem of Eq.~(\ref{EQ-GEV-EQ})
has a potential difficulty,
because an explicit orthogonalization procedure is required
with  ${\rm O}(N^3)$ operation costs,  
so as to satisfy
the orthogonality relation of $\bm{y}_k^{\rm T} B \bm{y}_l = \delta_{kl}$.
The above potential difficulty appears commonly among the large-scale electronic state calculations.
A calculation code with ${\rm O}(N^3)$ operation costs is RSDFT \cite{RSDFT}, the winner of Gordon Bell Prize in 2011.
The method is based on first principles and real-space mesh grid and
was used with up to $N=10^5$ atoms on the K computer.
Although the above paper is a fascinating progress, 
the present target is far beyond the computational limit.

\subsection{A novel concept for large-scale calculations \label{SEC-BACKGROUND-ON}}

A novel concept for large-scale calculations 
was proposed by Walter Kohn, a winner of the Nobel Prize in Chemistry at 1998.
His paper in 1996 shows that the above potential difficulty in electronic state calculation can be avoided, when the theory is not based on an eigenvalue problem 
and the formulation is free from the orthogonalization procedure
\cite{KOHN-1996}.
The concept realizes \lq order-$N$' methods
\cite{HOSHI-MARNOLDI, SIESTA, ONETEP, OZAKI-2006, CONQUEST},
in which the computational cost is ${\rm O}(N)$ or proportional to the number of atoms $N$.

Here the concept \cite{KOHN-1996} is  briefly explained.
The theory focuses on a physical quantity defined as
\begin{eqnarray}
\langle X \rangle \equiv \sum_k \, f(\lambda_k) \, \bm{y}_k^{\rm T} X \bm{y}_k,
\label{EQ-PHYS-X}
\end{eqnarray}
with a given sparse real-symmetric matrix $X$.
Equation (\ref{EQ-PHYS-X}) is found in elementary textbooks
of electronic state calculations.
The function of $f(\lambda)$ is a weight function, called Fermi function, and is defined as
\begin{eqnarray}
f(\lambda) \equiv \left\{ 1 + \exp(\frac{\lambda-\mu}{\tau}) \right\}^{-1}.
\label{EQ-FD-FUNC}
\end{eqnarray}
The weight function is a \lq smoothed' step function 
with a smoothing parameter $\tau(>0)$, 
because the Heaviside step function will appear in the limiting case of $\tau \rightarrow +0$
($f(\lambda)=1 (\lambda < \mu)$ and $f(\lambda)=0 (\lambda > \mu)$).
The smoothing parameter $\tau$ indicates the temperature of electrons.
The parameter $\mu$ is the chemical potential and the value should be determined, 
so as to reproduce the number of electrons in the material.
The case in $X=A$, for example, gives the electronic energy
\begin{eqnarray}
 \langle A \rangle  \equiv \sum_k f(\lambda_k) \, \lambda_k.
 \label{EQ-ELEC-ENE}
\end{eqnarray}
%

A quantity in Eq.(\ref{EQ-PHYS-X}) is transformed into the trace form of
\begin{eqnarray}
\langle X \rangle = {\rm Tr} [\rho X] = \sum_{i,j} \rho_{ji} X_{ij}
\label{TRACE-X}
\end{eqnarray}
with the density matrix
\begin{eqnarray}
\rho \equiv \sum_k \,  f(\varepsilon_k) \, \bm{y}_k \, \bm{y}_k^{\rm T}.
\label{DEF-DM}
\end{eqnarray}

The order-$N$ property can appear, since the matrix $X$ is sparse;
A density matrix element $\rho_{ji}$ is {\it not} required when $X_{ij} = 0$,
because the element $\rho_{ji}$ does not contribute to the physical quantity
of Eq.~(\ref{TRACE-X}), 
even if its value is nonzero ($\rho_{ji} \ne 0$).
Consequently, the number of the required density matrix elements $\rho_{ji}$ is ${\rm O}(N)$.
The above fact is called \lq quantum locality' or
\lq nearsightedness principle' \cite{KOHN-1996}.
%


The above formulation has a highly parallelizable mathematical structure
and the original problem is decomposed mathematically into parallel subproblems.
The trace in Eq.~(\ref{TRACE-X}) can be decomposed as
\begin{eqnarray}
\langle X \rangle = {\rm Tr} [\rho X] = \sum_j^M \bm{e}_j^{\rm T} \rho X \bm{e}_j,
\label{TRACE-X-DECOMP}
\end{eqnarray}
with the $j$-th unit vector of
$\bm{e}_j \equiv (0,0,0,...,1_j,0,0,...,0)^{\rm T}$.
Here the quantity of $\bm{e}_j^{\rm T} \rho X \bm{e}_j$ is called \lq projected physical quantity', because the quantity is defined by the projection onto the vector of $\bm{e}_j$.
The essence of the parallelism is the fact that the projected physical quantity of $\bm{e}_j^{\rm T} \rho X \bm{e}_j$ is calculated almost independently among different indices of $j$.

An important application is quantum molecular dynamics simulation,
in which an electron is treated as a quantum mechanical wave, 
while an atom (a nucleus) is treated as a classical particle
in Newtonian equation of motion
\begin{eqnarray}
M_I \frac{d^2 \bm{R}_I}{dt^2} = \bm{F}_I.
\end{eqnarray}
Here, $M_I$ and $\bm{R}_I$ are the mass and the position 
of the $I$-th atom and $\bm{F}_I$ is the force on the $I$-th atom. 
Other variables, such as the electronic charge on each atom $\{ q_I \}_I$, 
can be also calculated.
The force and charge on each atom can be calculated 
in the trace form of Eq.~(\ref{TRACE-X}).

\subsection{Physical origin of the matrices \label{SEC-LOCALITY}}

The matrices of $A$ and $B$ are sparse and 
their physical origin is found in
Ref.~\cite{HOSHI-MARNOLDI} and reference therein.
In short, the calculations in the present paper are formulated
by a first-principle-based modeled (transferable tight-binding) theory.
An electronic wavefunction $\phi_k(\bm{r})$ is expressed 
by an eigenvector of $\bm{y}_k \equiv (y_{1k}, y_{2k}, ..., y_{Mk})^{\rm T}$, as
$\phi_k(\bm{r}) = \sum_j y_{jk} \chi_j(\bm{r})$
with the given basis functions of $\{ \chi_j(\bm{r}) \}_j$ called atomic orbitals.
A basis function is localized in real space 
and its localization center is 
the position of one atom. %
A matrix element of $A_{ij}$ or $B_{ij}$ represents 
the quantum (wave) interaction of electrons on the $i$-th and $j$-th bases.
The basis index, $i$ or $j$, is the composite indices of the atom index $I$ or $J$ that distinguishes the localization center
and another index, $\alpha$ or $\beta$, called orbital index  
that distinguishes the shape of the function 
($i \Leftrightarrow (I, \alpha), j \Leftrightarrow (J, \beta)$).
An element of the matrices $A$ and $B$ can be expressed by the four indices as 
$A_{I\alpha;J\beta}$ and $B_{I\alpha;J\beta}$, respectively.
The matrices are sparse, 
because the elements decays quickly 
($|A_{I\alpha;J\beta}|, |B_{I\alpha;J\beta}| \rightarrow 0$) 
as the function of the distance 
between the $I$-th and $J$-th atoms ($r_{IJ}$).
In the present simulation,  a cutoff distance $r_{\rm cut}$ was introduced
so that a matrix element, $A_{I\alpha;J\beta}$ or $B_{I\alpha;J\beta}$, is ignored
in the cases of $r_{IJ} > r_{\rm cut}$.
Among the present benchmarks, 
the cutoff distance $r_{\rm cut}$ is set to be $r_{\rm cut}=5$ au ($\approx$ 0.2646nm) for diamond crystal and $r_{\rm cut}=10$ au ($\approx$ 0.5292nm) for condensed polymers.
A longer cutoff distance is used for condensed polymers, 
so as to include the interaction between polymers.

The number of orbitals on one atom can be different among atom species. 
The simulated materials in the present paper consists in hydrogen (H) and carbon (C) atoms. 
One (s-type) orbital is prepared at each hydrogen (H) atom, and
four (s-, p$_x$-, p$_y$-, p$_z$-types) atomic orbitals at each carbon (C) atom.
A material with $N_{\rm H}$ hydrogen atoms and $N_{\rm C}$ carbon atoms
gives the matrices of $A$ and $B$ with the size of $M = N_{\rm H} + 4N_{\rm C}$.

%
%
%

\begin{figure}[t]
\begin{center}
\unitlength=1mm
\includegraphics[width=8.5cm]{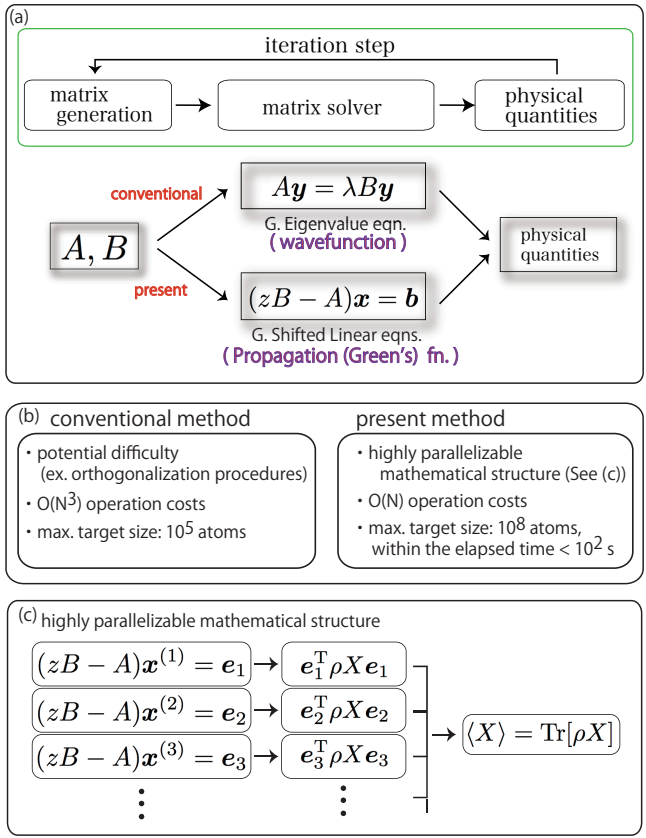}
\caption{
(a) The ground design of scalable algorithm.
(b) Comparison between the conventional and
present method.
(c) Illustration of the highly parallelizable mathematical structure.
\label{FIG-GROUND-DESIGN}
}
\end{center}
\end{figure}

\section{Algorithm \label{SEC-ALGO}}

\subsection{Ground design}

The ground design of the present scalable algorithm 
~\cite{HOSHI-MARNOLDI} 
is shown in Fig.~\ref{FIG-GROUND-DESIGN}(a).
The comparison between the conventional and present methods
is summarized in Fig.~\ref{FIG-GROUND-DESIGN}(b).
The method is based not on the eigenvalue problem of Eq.~(\ref{EQ-GEV-EQ}) 
but on the set of linear equations in the form of
\begin{eqnarray}
 ( z B -A ) \bm{x} = \bm{b}.
 \label{EQ-SHIFT-EQ}
\end{eqnarray}
Here, 
$z$ is a complex energy value.
The vector $\bm{b}$ is an input and the vector $\bm{x}$ is the solution vector.
A set of linear equations in the form of Eq.~(\ref{EQ-SHIFT-EQ})
with different energy values ($z=z_1, z_2, ... $) is
called generalized shifted linear equations.
The case in $B=I$ is  called shifted linear equations.
The use of Eq.~(\ref{EQ-SHIFT-EQ}) results in the Green's (propagation) function formalism,
since the solution $\bm{x}$ of  Eq.~(\ref{EQ-SHIFT-EQ}) is written formally as
\begin{eqnarray}
\bm{x} = G \bm{b}
\label{EQ-EIG-X=GB}
\end{eqnarray}
with the Green's function $G  \equiv (zB-A)^{-1}$.
The Green's function and the eigenvectors holds the relationship of
\begin{eqnarray}
G(z) = \sum_k^{M} \, \frac{ \bm{y}_k \, \bm{y}_k^{\rm T}}{z - \lambda_k}.
\label{EQ-EIG-GREEN}
\end{eqnarray}
The density matrix is also given by the Green's function as
\begin{eqnarray}
\rho =  \frac{-1}{\pi}  \int_{-\infty}^{\infty} \,  f(\varepsilon) \, {\rm Im} [G(\varepsilon + i 0)] \, d \varepsilon.
\label{EQ-DM-GREEN-FN}
\end{eqnarray}

The present method has
a highly parallelizable mathematical structure,
as illustrated in
Fig.~\ref{FIG-GROUND-DESIGN}(c), 
since the projected physical quantity of
$\bm{e}_j^{\rm T} \rho X \bm{e}_j$ in Eq.~(\ref{TRACE-X-DECOMP}) is
obtained from the generalized shifted linear equations of
\begin{eqnarray}
 ( z B -A ) \bm{x}^{(j)} = \bm{e}_j.
 \label{EQ-SHIFT-EQ-J}
\end{eqnarray}

\subsection{Krylov subspace solver}

The generalized shifted linear equations of Eq.~(\ref{EQ-SHIFT-EQ-J})
are solved on an iterative Krylov-subspace solver.
A Krylov subspace is defined 
as the linear space of
\begin{eqnarray}
K_\nu(Q;\bm{b}) \equiv {\rm span}[\bm{b}, Q \bm{b}, Q^2 \bm{b},..., Q^{\nu-1} \bm{b}],
\end{eqnarray}
with a given vector $\bm{b}$ and a given square matrix $Q$.
An example is Conjugate Gradient method and 
the subspace dimension of $\nu$ is the number of iterations.
Krylov-subspace methods with 
(generalized) shifted linear equations have been investigated 
in particular from 2000's, 
partially because the strategy is suitable to parallelism.
Since the solver algorithms are mathematical,
they are applicable to many scientific areas, such as,
QCD \cite{FROMMER-2003},
large-scale electronic state calculation 
\cite{HOSHI-MARNOLDI, TAKAYAMA-2006, SOGABE-2008-ETNA, TENG-2011-PRB, SOGABE-2012-JCP},
quantum many-body electron problem \cite{YAMAGEN-2008},
nuclear shell model problem \cite{MIZUSAKI-2010},
first-principle electronic excitation problem \cite{GIUSTINO-2010}, and
first-principle transport calculation \cite{IWASE-2015}.
In the present paper, 
the multiple Arnoldi solver \cite{HOSHI-MARNOLDI} is used,
in which 
Eq.~(\ref{EQ-SHIFT-EQ-J}) is solved within the direct sum of the two Krylov subspaces of
\begin{eqnarray}
{\cal L}_\nu(A,B;\bm{e}_j) \equiv  K_{\nu/2}(A;\bm{e}_j) \oplus K_{\nu/2}(A;B^{-1}\bm{e}_j)
\label{EQ-LS-MULTIPLE}
\end{eqnarray}
with an even number of $\nu$.
The number $\nu$ is typically, $\nu=30-300$ and the calculations in the present paper was carried out with $\nu=30$ as in the previous one \cite{HOSHI-MARNOLDI}.
The second term in the right hand side
of Eq.~(\ref{EQ-LS-MULTIPLE}) appears so as to satisfy several conservation
laws ~\cite{HOSHI-MARNOLDI}.
A reduced (small) $\nu \times \nu$ eigenvalue equation is solved and 
the solution vector is given by
\begin{eqnarray}
\bm{x}^{(j)} : = G^{(j)}(z) \bm{e}_j
\end{eqnarray}
with
\begin{eqnarray}
G^{(j)}(z) \equiv \sum_m^{\nu} \, \frac{ \bm{v}_m^{(j)} \, \bm{v}_m^{(j){\rm T}}} {z - \varepsilon_m^{(j)}}.
\label{EQ-EIG-GREEN-KRYLOV}
\end{eqnarray}
Here $\varepsilon_m^{(j)}$ and $\bm{v}_m^{(j)}$ is an eigenvalue 
and eigenvector of the reduced equation $(m=1,2,....\nu)$.
When the Green's function of $G$ in Eq.~(\ref{EQ-DM-GREEN-FN}) is replaced by $G^{(j)}(z)$ in Eq.~(\ref{EQ-EIG-GREEN-KRYLOV}),
the projected physical quantity with the index of $j$ is given by
\begin{eqnarray}
\bm{e}_j^{\rm T}  \rho X \bm{e}_j &:=&
\frac{-1}{\pi}  \int_{-\infty}^{\infty} \,  f(\varepsilon) \, {\rm Im} [\bm{e}_j^{\rm T}
G^{(j)}(\varepsilon \! + \! i 0)X \bm{e}_j] \, d \varepsilon  \nonumber \\
& =& \sum_m^{\nu} f(\varepsilon_m^{(j)})
\bm{e}_j^{\rm T} \bm{v}_m^{(j)} \, \bm{v}_m^{(j){\rm T}} X \bm{e}_j.
\label{EQ-PROJ-PHYS-QUANT-KRYLOV}
\end{eqnarray}
An advantage of the method is that 
the energy integration is carried out analytically
as in Eq.(\ref{EQ-PROJ-PHYS-QUANT-KRYLOV}).
Equation (\ref{EQ-PROJ-PHYS-QUANT-KRYLOV}) will be exact, 
if the subspace dimension of $\nu$ increases to the original matrix dimension ($\nu = M$).
As an additional technique in large-scale calculations, 
the real-space projection technique \cite{HOSHI-MARNOLDI} was also used.
The radius of the spherical region is determined
with an input integer parameter $\kappa$, so that the region contains $\kappa$ atoms or more.
The same technique is used also for the overlap matrix $B$.
The value of $\kappa$ is set to $\kappa=100$ in the present paper
as in the previous one \cite{HOSHI-MARNOLDI}.
As results,
numerical problems in the form of Eq.~(\ref{EQ-SHIFT-EQ-J})
are solved with the matrix size of, typically, $M' = 200-400$
in the present paper.

\subsection{Implementation  \label{SEC-TECH}}

The code is written in Fortran 90
with the MPI/OpenMP hybrid parallelism.
According to the parallel scheme in Fig.~1(c),
the projected physical quantity of $\bm{e}_j^{\rm T} \rho X \bm{e}_j$ is 
calculated as single-thread or single-core calculations.
As explained in Sec.~\ref{SEC-LOCALITY}, 
the basis index $j$ is a composite suffix of the atom index $J$ and 
the orbital index $\beta$ ($j \Leftrightarrow (J,\beta)$)
In the code, 
the loop for the basis index $j$ is implemented
as the double loop that consists of  
the outer loop for the atom index $J$
and the inner loop for the orbital index $\beta$. 
Since the outer loop is parallelized both in MPI and OpenMP parallelism,
a meaningful parallel computation is possible,
when the number of atoms is larger than that of cores
$(N > n_{\rm core})$.
Several matrix elements of $A,B$ are generated redundantly among nodes,
so as to save inter-node communications.
The pure MPI parallelism is possible but consumes larger memory costs.

The communication among nodes is required, 
{\it only when} a summation is performed in the trace form of Eq.~(\ref{TRACE-X-DECOMP}), 
as shown in Fig.~\ref{FIG-GROUND-DESIGN}(c) 
\cite{HOSHI-MARNOLDI}. 
The summation is carried out hierarchically;
First, the summation is carried out on each node by OpenMP directives and 
then the summation is carried out between nodes by \verb|MPI_Allreduce()|.

\section{Benchmark  \label{SEC-BENCH}}

\subsection{Purpose and condition}

The benchmarks were carried out
so as to show an extreme strong scaling and 
a time-to-solution
qualified for a real research.
Our target value of the qualified time-to-solution is
$T_{\rm elaps}=10^2$s for the elapsed time per step
in a quantum molecular dynamics simulation,
because a dynamical simulation of $n_{\rm step}=10^3$ steps
can be executed within one day
($ T_{\rm elaps} n_{\rm step} = 10^5$s $\approx$ one day).
The calculations were carried out on the K computer 
which consists of 82,944 compute nodes and 
achieved the peak performance of 11.28PFLOPS.
Each CPU has eight cores and 
the interconnection between nodes 
is named \lq Tofu' which constructs 
physical six-dimensional mesh/torus network topology.
We used the \verb|MPI_Allreduce()| 
optimized on the K computer
\cite{ADACHI-2013-MPI}.

The calculations were executed in double precision
with the MPI/OpenMP hybrid parallelism.
The number of the MPI processes
is set to that of the compute nodes and
the number of the OpenMP threads
is set to be eight, the number of cores per compute node.
The jobs were executed
by specifying the three-dimensional
node geometry on the K computer
$(n_{\rm node} \equiv n_{\rm node}^{(x)} \times  n_{\rm node}^{(y)} \times n_{\rm node}^{(z)})$
for optimal performance or minimum hop count.
The number of used nodes (node geometry) is listed below;
$n_{\rm node}$ =
$2,592 (=12 \times 12 \times 18)$,
$5,184 (=12 \times 18 \times 24)$,
$10,368 (=18 \times 24 \times 24)$,
$20,736 (=24 \times 27 \times 32)$,
$41,472 (=27 \times 32 \times 48)$, and
$82,944 (=32 \times 48 \times 54$, the full system).

The benchmark were carried out
for disordered materials that appears in real research 
of ultra-flexible devices.
Condensed polymer systems of 
poly-(phenylene-ethynylene) (PPE)
were simulated. 
The three systems are called \lq P100', \lq P10' and \lq P1' and
contain $N$=101,606,400 ($\approx 10^8$ or 100M),
$N$=10,137,600 ($\approx$ 10M) and
$N$=1,228,800 ($\approx$1M) atoms,
respectively.
The periodic boundary condition is imposed.
The size of the periodic simulation box is
134 nm $\times$ 134 nm $\times$ 209 nm
for the \lq P100' system.
The simulations were carried out also for
the ideal diamond solid called \lq D100' that contains
$N=106,168,320$ ($\approx 10^8$ or 100M) atoms in the ideal periodicity,
so as to discuss the influence of the presence or absence of structural disorder.

Technical details are explained.
The initial atomic structures for the polymer systems
were generated in classical molecular dynamics simulations
by GROMACS
(\url{http://www.gromacs.org/}).
Classical simulations work faster but do not
treat electronic (quantum)  waves responsible for
the device property.
The recorded elapsed time was one 
for  a \lq snapshot' simulation,
an electronic state calculation of the given atomic structures,
which dominates
the elapsed time in molecular dynamics simulations.
A molecular dynamics simulation 
can not be carried out with $N=10^8$ atoms,
because 
the required memory size exceeds 
the limit of the K computer (16GB per node).
The present snapshot calculation with $N=10^8$ atoms 
consumes 9 GB per node and 
a molecular dynamics simulation requires a larger memory size,
so as to store additional variables like velocity, force and so on.
The benchmark of  molecular dynamics simulation with $N=10^7$ atoms
will be discussed in the last paragraph of this section.

\begin{table}[b]
 \begin{center}
  \caption{The measured elapsed times $T_{\rm elaps}$ (sec)
  for ideal diamond solid with $10^8$ atoms (\lq D100') and
  condensed polymer systems with  $10^8$ atoms (\lq P100'),
  with $10^7$ atoms (\lq P10') and with $10^6$ atoms (\lq P1').
  The ideal or measured speed-up ratio is shown inside the parenthesis.
  \label{TABLE-BENCH-DATA}}
  \begin{tabular}{|c|c|c|c|c|} \hline
   $n_{\rm node}$ & D100 & P100 & P10 & P1 \\ \hline
 2,592 (1)  & 1001.4 (1)     &  741.1 (1)     & 81.4 (1)     & 10.3 (1)    \\ \hline
 5,184 (2)  &  502.2 (1.99)  &  378.5 (1.96)  & 43.7 (1.86)  &  5.95 (1.73) \\ \hline
10,368 (4)  &  252.6 (3.96)  &  195.2 (3.80)  & 24.3 (3.35)  &  3.28 (3.14) \\ \hline
20,736 (8)  &  127.9 (7.83)  &  103.0 (7.19)  & 11.4 (7.14)  &  1.96 (5.26) \\ \hline
41,472 (16) &   65.6 (15.3)  &   57.1 (13.0)  & 6.32 (12.9)  &  1.25 (8.23) \\ \hline
82,944 (32) &   34.1 (29.4)  &   30.9 (24.0)  & 3.60 (22.6)  &  0.84 (12.2) \\ \hline

   \end{tabular}
  \end{center}
\end{table}

\begin{figure}[t]
\begin{center}
\unitlength=1mm
\includegraphics[width=5cm]{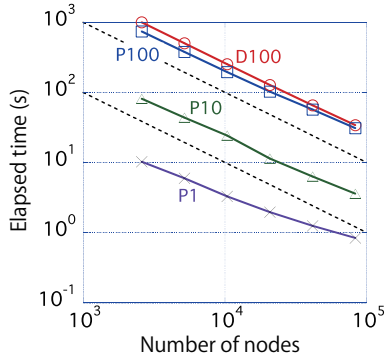}
\caption{Strong scaling benchmarks 
for ideal diamond solid with $10^8$ atoms (\lq D100') and
the condensed polymer systems with  $10^8$ atoms (\lq P100'),
with $10^7$ atoms (\lq P10') and with $10^6$ atoms (\lq P1').
Dashed lines are drawn for ideal scaling.
\label{FIG-BENCHMARK-STRONG}
}
\end{center}
\end{figure}

\subsection{Result}

The measured elapsed time is summarized in
Table \ref{TABLE-BENCH-DATA}.
Here the parallel efficiency ratio $\alpha$ is defined by
\begin{eqnarray}
\alpha \equiv
( T_{\rm elaps}(n_0)/T_{\rm elaps}(n_{\rm node})) /
(n_{\rm node}/ n_0)
\end{eqnarray}
with $n_0 \equiv 2,592$.
For example,
the parallel efficiency ratio $\alpha$
with $10^8$ atoms and the maximum number of nodes ($n_{\rm node}=82,944$)
is
$\alpha =0.92$ for \lq D100'
and
$\alpha =0.75$ for \lq P100'.

Figure ~\ref{FIG-BENCHMARK-STRONG} shows 
the strong scaling property 
by plotting  the data of Table \ref{TABLE-BENCH-DATA}.
In all the cases, 
the elapsed time $T_{\rm elaps}$
decreases monotonically as the function of the number of used nodes.
The order-$N$ property ($T_{\rm elaps} \propto N$) is also found.
For example,  
the time of \lq P100' is ten times larger that of \lq P10' 
with $n_{\rm node} = 2,592$.
As a rough estimation from Fig.~\ref{FIG-BENCHMARK-STRONG},
the target time-to-solution
of $T_{\rm elaps} \approx 10^2$s is fulfilled
by $n_{\rm node} \approx 2 \times 10^4$ and $2 \times 10^3$
for the condensed polymer systems with $N=10^8$ and $10^7$ atoms,
respectively.
The two cases conclude commonly that 
the qualified time-to-solution is fulfilled, 
when
the number of atoms per node
is approximately $5 \times 10^2$
$(N/n_{\rm node} \approx 5 \times 10^2)$.
The above statement can be interpreted
as the weak-scaling property.

\begin{table}[b]
 \begin{center}
  \caption{Communication time $T_{\rm comm}$  and
  barrier time $T_{\rm barr}$ of the elapsed time $T_{\rm elaps}$.
  See the caption of Table \ref{TABLE-BENCH-DATA} for notations.
  The values are listed as $T_{\rm comm}$ / $T_{\rm barr}$ (sec).
  \label{TABLE-BENCH-DATA2}}
  \begin{tabular}{|c|c|c|c|c|} \hline
   $n_{\rm node}$ & D100          & P100          & P10           & P1             \\ \hline
 2,592            & 1.04 /  7.16  & 1.60 / 28.14  & 0.382 / 6.69  & 0.0617 / 2.68  \\ \hline
 5,184            & 1.04 /  3.34  & 1.60 / 20.75  & 0.378 / 4.86  & 0.0689 / 1.85  \\ \hline
10,368            & 1.05 /  2.22  & 1.61 / 14.97  & 0.384 / 4.34  & 0.0734 / 1.13  \\ \hline
20,736            & 1.05 /  1.34  & 1.61 /  9.05  & 0.218 / 3.31  & 0.0712 / 0.674 \\ \hline
41,472            & 1.06 /  1.03  & 1.64 /  6.87  & 0.215 / 2.18  & 0.0727 / 0.409 \\ \hline
82,944            & 1.06 / 0.485  & 1.65 /  5.96  & 0.218 / 1.28  & 0.0613 / 0.227 \\ \hline
   \end{tabular}
  \end{center}
\end{table}

\begin{figure}[t]
\begin{center}
\unitlength=1mm
\includegraphics[width=6.5cm]{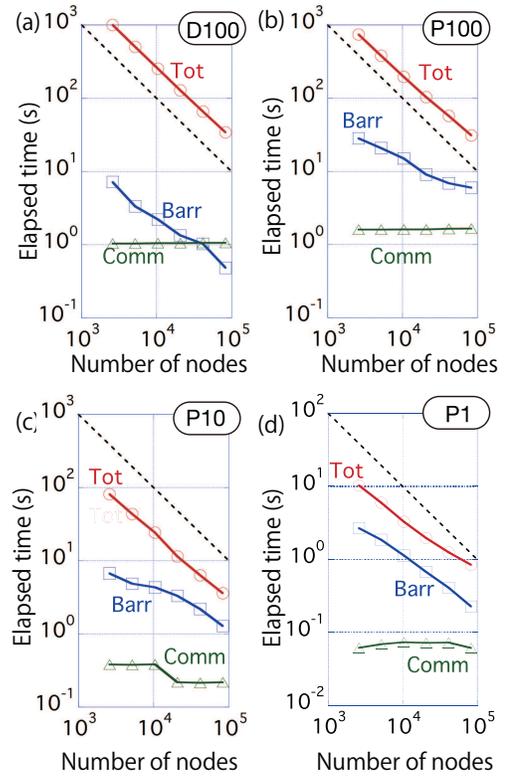}
\caption{Details  of the elapsed time.
The total elapsed time $T_{\rm elaps}$ (Tot),
the communication time $T_{\rm comm}$ (Comm) and
the barrier time $T_{\rm barr}$ (Barr) are plotted.
See Fig.~\ref{FIG-BENCHMARK-STRONG} for notations.
\label{FIG-BENCHMARK-ANALYSIS}
}
\end{center}
\end{figure}

\subsection{Analysis and discussion}

Table \ref{TABLE-BENCH-DATA2} shows 
the measured communication and barrier times.
In the simulations, we recorded not only
the total elapsed time $T_{\rm elaps}$,
but also
the accumulated MPI communication time $T_{\rm comm}$
and the accumulated barrier time $T_{\rm barr}$ on all nodes.
The barrier time includes the time to wait for other processors.
The communication time $T_{\rm comm}$ is consumed
by inter-node data communications,
while the barrier time $T_{\rm barr}$ appears
from a load imbalance among nodes.

Figure \ref{FIG-BENCHMARK-ANALYSIS} plots
the data in Tables  \ref{TABLE-BENCH-DATA} and \ref{TABLE-BENCH-DATA2}.
Two points are discussed;
(i) The communication time is not serious among all the cases. 
(ii) When the cases of \lq D100' and \lq P100' are compared,
the ratio of the barrier time is much larger than in \lq P100'.
In the full system calculation ($n_{\rm node}=82,944$),  for example,
the ratio is $T_{\rm barr} / T_{\rm elaps} \approx 0.19$ in \lq P100'
and is $\approx 0.014$ in \lq D100'. 
We should recall that 
the \lq D100' case is an ideal system without structural disorder and 
all the subproblems in Fig.~\ref{FIG-GROUND-DESIGN}(c) are equivalent. 
On the other hand, 
the load imbalance among nodes appears in \lq P100',
because of the structural disorder. 
The same conclusion holds on the \lq P10' and \lq P1' cases.
A method for better load balance is a future (not urgent) issue of the present code.

\begin{figure}[t]
\begin{center}
\unitlength=1mm
\includegraphics[width=5cm]{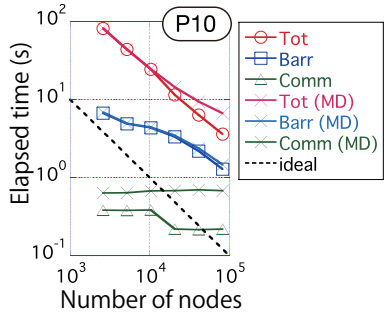}
\caption{Details  of the elapsed time for the MD simulation 
in the \lq P10' case. 
The total elapsed time $T_{\rm elaps}^{\rm (MD)}$ (Tot(MD)),
the barrier time $T_{\rm barr}^{\rm (MD)}$ (Barr(MD), and 
the communication time $T_{\rm comm}^{\rm (MD)}$ (Comm(MD)) are plotted per MD step
in the same manner of Fig.~\ref{FIG-BENCHMARK-ANALYSIS}(c).
The data for the electronic state calculation 
(Tot, Barr, Comm) are also plotted for comparison.
\label{FIG-BENCHMARK-MD}
}
\end{center}
\end{figure}

To end up this section,
two comments are addressed;
(I) 
The further tuning should be focused mainly on single-core calculations, 
since the most routines are executed  
as single-core calculations  
as in  Fig.\ref{FIG-GROUND-DESIGN}(c). 
The profiler reported that the performance  
is 2.3 \%  of the peak 
for the \lq P100' case with $n_{\rm node} = 82,944$
in Table \ref{TABLE-BENCH-DATA}. 
The severest limitation in the present calculations
is the memory size of the K computer (16GB per node) and 
the present code was written in the memory-saving style,
in which the memory cost should be minimized
and the time cost is sometimes sacrificed.  
Since the situation can differ among materials and/or architectures,
a possible way is to add another workflow in the time-saving style.
The routines can be classified into 
those for the generation of matrix elements and 
and for the Krylov subspace solver 
as in Fig.\ref{FIG-GROUND-DESIGN}(a). 
The matrix-vector multiplication gives
a large fraction of the total elapsed time, 
as usual in a Krylov-subspace solver, 
and a typical fraction is $21$ \%
among  the present condensed polymer systems.
The result suggests that the matrix generation part
gives a larger fraction. 
(II) Fig.~\ref{FIG-BENCHMARK-MD} shows 
the benchmark of molecular dynamics simulation 
for the \lq P10'  case,
the possible maximum size
(See the first paragraph of the present section),
in the same manner of Fig.~\ref{FIG-BENCHMARK-ANALYSIS}(c).
For example, 
the elapsed time per molecular dynamics time step is 
$T_{\rm elaps}^{\rm (MD)}$ = 81.8 sec or  6.62 sec 
in $n_{\rm node} =$2,592 or 82,944, respectively.
For comparison, 
Fig.~\ref{FIG-BENCHMARK-MD} also shows 
the data in Fig.~\ref{FIG-BENCHMARK-ANALYSIS}(c), 
the data with the electronic structure calculation part.
The elapsed time is much smaller than 
the target time-to-solution ($10^2$s) 
and the method is qualified well for a real research. 
We found, however, that non-negligible time costs 
appear in the total elapsed time ($T_{\rm elaps}^{\rm (MD)}$)
among the cases with $n_{\rm node} > 2 \times 10^4$,
because of the additional routine for MD simulation. 
Now we are tuning the code for faster MD simulations.

\section{Application in real material research \label{SEC-SCIENCE}}

This section is devoted to the application study of the present method 
to a condensed polymer system,
so as to show how a real research works well with $N=10^8$ atoms
by distributed computing. 
As an application study with a smaller system, 
a molecular dynamics simulation with $N=10^5$ atoms
was carried out with $10^4$ cores and
the elapsed time is 10 hours for 5,000 iteration steps 
\cite{HOSHI-GREECE2}.
Such a dynamical simulation is 
impractical  with $N=10^8$ atoms at the present day and
this section indicates a part of
the possible future research.

Here, the condensed organic polymer system of \lq P100' was used.
The research is motivated by 
an academic-industrial collaboration 
with Sumitomo Chemical Co., Ltd.
\cite{HOSHI-MARNOLDI, HOSHI-2014-JPSCP, HOSHI-2014-POS}.
Organic material gives the foundation of ultra-flexible (wearable) devices, 
key devices of next-generation IoT products, 
such as display, sensor and battery. 
A recent example is \lq e-skin' \cite{SOMEYA-2016}.
The material is ultra-flexible (soft) and disordered in structure and 
the thickness of devices is typically $10^3$ nm and 
100-nm-scale simulations are crucial. 

An important HPC issue is that
the distributed data structure should be preserved 
throughout the whole research; 
Since the simulation data is huge and distributed, 
we cannot gather them into one node. 
Here we will show that the post-simulation data analysis 
works well for distributed data.

\begin{figure*}[t]
\begin{center}
\unitlength=1mm
\includegraphics[width=18cm]{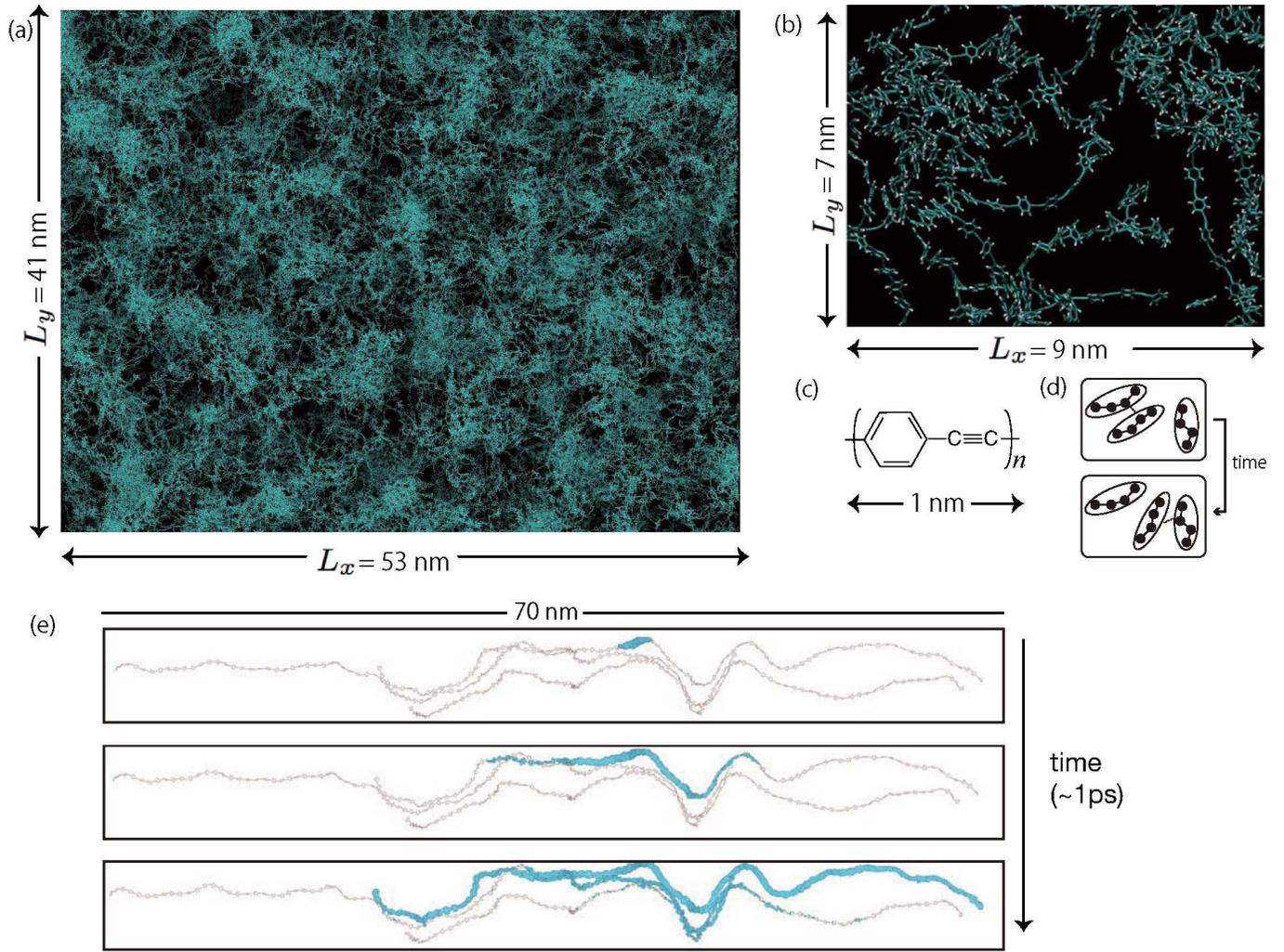}
\caption{
A real material research for a condensed polymer system
(PPE) with $10^8$-atoms.
(a) (b) Visualization of partial regions.
The whole system has the periodic cell lengths of (265nm, 206nm,  239nm).
(c) The unit structure of the polymer (PPE).
(d) Schematic figures of a dynamically changed local network of connected polymers.
(e) Quantum wave dynamics simulation of electrons with a local network of 
three connected polymers.
The charge density of $q(\bm{r}, t) \equiv |\Psi(\bm{r}, t)|^2$ is drawn 
in the upper ($t=0$), middle($t=50$fs) and lower ($t=948$fs) panels.
\label{FIG-CONNECTIVITY}
}
\end{center}
\end{figure*}

Figures \ref{FIG-CONNECTIVITY}(a)(b) show
partial regions of the system and one 
can observe that the structure is fairly disordered.
The molecular structure for a polymer unit is shown 
in Fig.~\ref{FIG-CONNECTIVITY}(c).
In general, electronic wavefunctions are localized in a disordered structure.
The electrical current can propagate among polymers
that are \lq connected' locally by 
characteristic ($\pi$-type) electronic waves. 
We should investigate, therefore, the network of connected polymers.

\subsection{Network analysis of electronic wavefunctions}

A large-scale post-simulation data analysis was carried out
so as to characterize the propagation of electronic wave
in the disordered structure.
A speculated propagation mechanism is shown 
schematically in Fig.~\ref{FIG-CONNECTIVITY}(d).
Three polymers are drawn and
atoms are depicted as filled circles.
The figures include
a small local network that
consists of two polymers connected by a dashed line.
Electron can propagate along connected polymers.
Since the network structure is dynamically changed,
as schematically shown in Fig.~\ref{FIG-CONNECTIVITY}(d),
electron can propagate through the whole material.

The purpose of the analysis 
is to detect local polymer networks in which 
electronic wave can propagate.
The analysis was carried out
with the Green's function $G$ obtained 
by the parallel order-$N$ simulation,
as follows;
Stage I:
The present parallel simulation gives
a \lq connectivity' matrix of $C_{IJ}$
\begin{eqnarray}
C_{IJ}
& \equiv & \sum_\alpha \sum_\beta  \rho_{I\alpha;J\beta} H_{J\beta;I\alpha}
\end{eqnarray}
where $I,J$ are the atom indices.
The connectivity matrix is called
integrated crystal orbital Hamiltonian population (ICOHP)
among physics papers \cite{COHP-1993, HOSHI-2013-JPSJ}.
The quantity is a partial sum of the electronic energy $\langle H \rangle$
in Eq.~(\ref{EQ-ELEC-ENE})
($\langle H \rangle = \sum_{IJ} C_{IJ}$).
Since the matrix elements are calculated
always during the parallel simulation,
the elements can be obtained independently among nodes,
without any additional operation or communication cost.
A matrix element $C_{IJ}$ has a physical meaning of a local
bonding energy between the $I$-th and $J$-th atoms;
If the value of $|C_{IJ}|$ is significantly large,
the two atoms are \lq connected' by electronic wave.
Stage II:
Since every atom belongs to one of polymers, the connectivity matrix for polymers is defined by
\begin{eqnarray}
C^{(\rm poly)}_{PQ} \equiv \sum_I^{\in P} \sum_J^{\in Q} C_{IJ},
\end{eqnarray}
where the summation of $\sum_I^{\in P}$, for example, means the summation among the atoms that belong to the $P$-th polymer.
If an element $C^{(\rm poly)}_{PQ}$ shows a meaningful non-zero value,
the $P$-th and $Q$-th polymers are connected by electronic wave.
The matrix $C^{(\rm poly)}$ is sparse.
The dimension of $C^{(\rm poly)}$
is equal to the number of polymers $N^{\rm (poly)} =83,349$
and is much smaller than that of $C$ ($N=10^8$).
Stage III:
As a coarse grained analysis,
the eigenvalue equation of
$C^{(\rm poly)} \bm{z} = \lambda \bm{z}$ in the matrix dimension of $N^{\rm (poly)}$
was solved by the parallel eigenvalue solver 
\cite{IMACHI-2016-EIGENKERNEL}.
As results,
several eigenvectors $\bm{z}$ have several non-zero elements,
which means the presence of small local networks with several connected polymers.

The network analysis reveals
that the condensed polymer system has
small networks that consist of several polymers,
as illustrated in Fig.~\ref{FIG-CONNECTIVITY}(d).

\subsection{Quantum wave dynamics simulation for device property\label{SEC-RESULT-QWP}}

Quantum wave (wavepacket) dynamics simulation
\cite{HOSHI-GREECE2} was carried out for device property,
so as to confirm that the above network analysis is fruitful
or that an electronic wave can propagate in
the small polymer networks detected in the above analysis.
In the wave dynamics simulation, 
an electronic wave $\Psi(\bm{r},t)$, a complex scalar vector, 
propagates dynamically under 
a Schr\"odinger-type equation of $i \partial_t \Psi = H \Psi$ 
with an effective Hamiltonian (matrix) $H$.
See Ref.~\cite{HOSHI-GREECE2} and the references therein for details.
The atom positions also change dynamically.
Since the norm $q(\bm{r},t) \equiv |\Psi(\bm{r},t)|^2$ is the charge distribution, 
its dynamics gives the charge propagation or 
the (non-stationary) electrical current.
Figure \ref{FIG-CONNECTIVITY}(e) shows
a typical dynamical simulation for approximately 1 ps.
The simulation consumes six hours with 128 nodes.
The result shows that
the electronic wave propagates within a polymer first,
and later propagates into other polymers, as expected.
The method for large-scale wave dynamics simulation 
is under way.

\section{Conclusion \label{SEC-CONCLUSION}}

A novel linear algebraic algorithm realizes
$10^8$ atom or 100-nm-scale quantum material simulations
with an extreme scalability and a qualified time-to-solution
on the full system of the K computer.
The mathematical foundation is generalized shifted linear equations, 
instead of conventional generalized eigenvalue equations
and has a highly parallelizable mathematical structure.
The method was demonstrated 
in a real material research for next-generation 
IoT products. 
The present paper shows that 
an innovative scalable algorithm for a real research
can appear by the co-design 
among application, algorithm and architecture.

\section*{Acknowledgment}

Several atomic structure data were provided from Masaya Ishida (Sumitomo Chemical Co., Ltd.).
Part of the results is obtained by using the K computer at the RIKEN Advanced Institute for Computational Science (Proposal number hp150144, hp150281, hp160066, hp160222).
This research is partially supported  by Japan Science and Technology Agency, Core Research for Evolutional Science and Technology (JST-CREST) in the research area of \lq Development of system software technologies for post-peta scale high performance computing'.
This research is partially supported also by Grant-in-Aid for Scientific Research (KAKENHI Nos. 26400318 and 26286087, 16KT0016) from the Ministry of Education, Culture, Sports, Science and Technology (MEXT) of Japan. A part of the research is based on the collaboration with Priority Issue (Creation of new functional devices and high-performance materials to support next-generation industries) to be tackled by using Post \lq K' Computer, MEXT, Japan.

\balance

\end{document}